\documentclass[twocolumn,showpacs]{revtex4}

\usepackage{graphicx}%
\usepackage{dcolumn}
\usepackage{amsmath}

\makeatletter
\def\btt#1{\texttt{\@backslashchar#1}}%
\DeclareRobustCommand\bblash{\btt{\@backslashchar}}%
\makeatother

\begin{document}

\preprint{MgB2 1/T1}

\title[Short Title]{Evidence for Strong-coupling $S$-wave Superconductivity in MgB$_2$ :$^{11}$B NMR Study}

\author{H.~Kotegawa$^{1}$, K.~Ishida$^{1}$, Y.~Kitaoka$^{1,3}$, T.~Muranaka$^{2}$, and J.~Akimitsu$^{2,3}$}
\affiliation{
$^1$Department of Physical Science, Graduate School of Engineering Science, Osaka University, Toyonaka, Osaka 560-8531, Japan\\
$^{2}$Department of Physics, Aoyama-Gakuin University, Setagaya-ku, Tokyo 157-8572, Japan.\\
$^{3}$Core Research for Evolutional Science and Technology (CREST) of the Japan Science and Technology Corporation (JST)
}

\date{Februray 15, 2001}

\begin{abstract}
We have investigated a gap structure in a newly-discovered superconductor, MgB$_2$ through the measurement of $^{11}$B nuclear spin-lattice relaxation rate, $^{11}(1/T_1)$.  $^{11}(1/T_1)$ is proportional to the temperature ($T$) in the normal state, and decreases exponentially in the superconducting (SC) state, revealing a tiny coherence peak just below $T_c$. The $T$ dependence of $1/T_1$ in the SC state can be accounted for by an $s$-wave SC model with a large gap size of 2$\Delta /k_BT_c \sim$ 5 which suggests to be in a strong-coupling regime. 
\end{abstract}

                       
\maketitle

Quite recently, Akimitsu and co-workers have discovered a new superconducting (SC) material MgB$_2$ that reveals a remarkably high SC transition temperature of $T_c\sim 40$ K \cite{Akimitsu}.
MgB$_2$ crystallizes in the hexagonal AlB$_2$-type structure, consisting of alternating hexagonal layers of Mg atoms and graphite-like layers of B atoms. 
The discovery of superconductivity with a relatively high value of $T_c$ in this compound gives a new impact in the solid state physics, since it may give other possibility for  finding the high-$T_c$ superconductivity in some binary intermetallic compounds besides  cuprates and C$_{60}$-based compounds. It may be promising to discover new compounds with a high-$T_c$ value exceeding the liq.-N$_2$ temperature other than cuprates.
Soon after the discovery, Bud'ko and co-workers reported that $T_c$ increases from 39.2 K for Mg$^{11}$B$_2$ to 40.2 K for Mg$^{10}$B$_2$, giving a clear indication that electron-phonon interactions are playing an important role \cite{Bud'ko}.  However, the observed high $T_c$ value of this compound seems to be either beyond or at a limitation for the phonon-mediated superconductivity that was predicted theoretically several decade ago\cite{McMillan}. Therefore we speculate on that a new exotic mechanism might be possible for the occurrence of the high-$T_c$ superconductivity in MgB$_2$. 
In order to gain an insight into a possible mechanism, it is quite important to understand its pairing symmetry and a SC-gap structure.

In this letter, we report $^{11}$B-NMR study that has shed light on the above issues.  Spin-lattice relaxation rate ($1/T_1$) of $^{11}$B has been measured in a temperature ($T$) range of 15 - 260 K.
We have found a $T_1T$=constant behavior in the normal state, and an exponential decrease of $1/T_1$ that was accompanied with a tiny coherence peak in the SC state. 
From the $T$ dependence of $1/T_1$ in the SC state that is consistent with an $s$-wave model with a larger isotropic gap of $2\Delta/k_BT_c\sim$ 5 than the BCS value of 3.5, the superconductivity in MgB$_2$ is concluded to be in a 
strong coupling regime and suggested to be mediated by strong-electron phonon 
interactions. 

A polycrystalline sample of MgB$_2$ was prepared as in reference\cite{Nagamatsu}.  
Electric resistivity and DC magnetization show the SC transition at 39 K.
The bulk sample without powdering has been used for NMR measurements to avoid some crystal defects if any.
$T_1$ was measured by monitoring the nuclear magnetization after a saturation rf-pulse under an external field $H=$13.5 kOe at a frequency $f$=18.5 MHz in the normal state and $H$=44.2 kOe at $f$=60.4 MHz in the SC state.
$T_c$=29 K under $H$=44.2 kOe was determined from an ac-susceptibility that was measured using an "in-situ" NMR coil, as shown in an inset of Fig.1.
The result is in good agreement with the references reported recently\cite{Finnemore,Takano}.

$^{11}$B-NMR spectrum has a simple shape with a full width at the half maximum of $\sim$ 10 Oe.  
Its resonance peak does not show any $T$ dependence in the normal state within an experimental error of 2 Oe.
$T_1$ measured at the peak in spectrum was determined by fitting the relaxation function of the nuclear magnetization $m(t)$ to the following theoretical two-exponential form,\cite{Narath}
\begin{eqnarray}
\begin{array}{rcl}
m(t)&=&\frac{M(\infty)-M(t)}{M(\infty)} \\
    &=& \frac{1}{10}\exp\left(-\frac{t}{T_1}\right) +\frac{9}{10}\exp\left(-\frac{6t}{T_1}\right).
\end{array}
\end{eqnarray}
Here $M(t)$ is the nuclear magnetization at time $t$ after saturation pulses.
From good fits to this formula, $T_1$ was precisely determined over 
a measured $T$ range, although a short component of $T_1$ appears at low temperatures below 20 K, associated with the presence of vortex cores in the mixed state. 

Figure 1 shows the $T$ dependence of $1/T_1$ in the $T$ range of 15 - 260 K. 
In the normal state, $1/T_1$ obeys a $T_1T$=constant relation with $T_1T$ = 1.8$\times 10^2$ ($s \cdot$ K) down to $T_c$.
In the SC state, we have found a tiny coherence peak in $1/T_1$ just below $T_c$, followed by an exponential decrease of $1/T_1$. 

In general, $1/T_1$ in the SC state is related to the density of states (DOS) in the SC state, $N_s(E)$ as follow,
\begin{equation}  
\frac{1}{T_1} \propto \int_{0}^{\infty}{(N_s(E)^2+M(E)^2)f(E)(1-f(E))}dE
\end{equation}
where $M(E)$ and $f(E)$ are the so-called "anomalous" DOS arising from the coherence effect of scattering inherent to a spin-singlet SC state and the Fermi-distribution function, respectively \cite{MacLaughlin}.
Note $M(E)$ =0 in the case of spin-triplet pairing state. 
The $T$ dependence of $1/T_1$ is calculated by using a typical $s$-wave model where a phenomenological energy broadening function in $N_s(E)$ is assumed to be of a rectangle type with a width $2\delta$ and a height $1/2\delta$ as presented by Hebel \cite{Hebel}.
A calculation is shown by a solid curve in Fig.1.
The $T$ dependence of $1/T_1$ except just below $T_c$ can be well reproduced by the calculation that assumes the SC gap size of $2\Delta/k_{\rm B}T_{\rm c}\sim 5$ and $\delta/\Delta(0)\sim 1/3$.
It should be noted that the $T$ dependence of $1/T_1$ cannot be reproduced by a model of the spin-triplet Balian-Werthamer (BW) state with an isotropic gap\cite{BW}, therefore the spin-triplet state should be ruled out. 
Remarkable findings of the tiny coherence peak in $1/T_1$ just below $T_c$ that is followed by the exponential decrease below 0.8 $T_c$ provide convincing evidence for the spin-singlet $s$-wave pairing realized in MgB$_2$. 
The  SC-gap value larger than  the weak-coupling BCS value suggests that the superconductivity in MgB$_2$ is in a strong coupling regime.   
The strong electron-phonon coupling may give rise to some intense lifetime broadening of quasiparticles just below $T_c$, suppressing the coherence peak more substantially than a behavior expected from the calculation.
The $T$ dependence of $1/T_1$ in the SC state of MgB$_2$ is quite similar to that in  Chevrel-phase compound TlMo$_6$Se$_{7.5}$ with $T_c$ = 12.2 K \cite{Ohsugi}.
Therefore we conclude that the superconductivity in MgB$_2$ is of a strong-coupling $s$-wave type with the larger value $2\Delta/k_BT_c\sim5$ than the weak coupling BCS one.

Present results of $1/T_1$ in MgB$_2$ are in quite contrast with that in cuprates such as La$_{1.85}$Sr$_{0.15}$CuO$_4$ that reveals a comparable $T_c$ value of 38 K \cite{Ishida}.
The $1/T_1T$ in the cuprates shows a Curie-Weiss behavior in the normal state, which evidences that the superconductivity appears on the verge of the antiferromagnetism. 
$1/T_1$ in the SC state decreases sharply without the coherence peak just below $T_c$, followed by a $T^3$ dependence far below $T_c$. 
This evidences that the symmetry of the Cooper pairs is of a $d$-wave type with a line-node gap. 
Various experiments that were reported so far tell us  that the Cooper pairs are presumably mediated via magnetic interactions other than electron-phonon interactions.  
On the other hand, the $T$ dependence of $1/T_1$ in MgB$_2$ shows that magnetic correlations are absent in the normal state and that the SC state is characterized by the large isotropic gap that is consistent with the strong coupling $s$-wave superconductor. 
In this context, it is natural to consider that the mechanism of the superconductivity in MgB$_2$ is quite different from in cuprates. 
Rather the electron-phonon interactions may play an important role. 
We suggest that the high value of $T_c$ in MgB$_2$ may be due to the strong electron-phonon coupling via some phonon modes inherent to a light element of boron.

In conclusion, we report that the $1/T_1$ of $^{11}$B decreases exponentially in the SC state, revealing a tiny coherence peak just below $T_c$.
The $T$ dependence of $1/T_1$ in the SC state can be accounted for by a spin-singlet $s$-wave model where 2$\Delta /k_BT_c \sim$ 5 is substantially larger than the weak coupling BCS value of 3.5.  
It is suggested that the strong electron-phonon interaction may play an important role for the high temperature superconductivity in MgB$_2$. 

We are grateful for helpful discussion with G.-q. Zheng.
This work was supported by the COE Research (10CE2004) in Grant-in-Aid for Scientific Research from the Ministry of Education, Sport, Science, and Culture of Japan.

\begin{figure}[htbp]
\caption[]{\protect $T$ dependence of $^{11}(1/T_1)$ in MgB$_2$. Closed and open circles correspond to $1/T_1$ measured in the external field of 44.2 kOe and 18.5 kOe, respectively. A solid curve in the SC state shows the calculation using the typical $s$-wave model (see text)\cite{Hebel}. The inset shows $T$ dependence of $ac$-susceptibility in 44.2 kOe measured by an {\it in-situ} NMR coil. The arrows show an onset of the superconductivity under 44.2 kOe.
}
\end{figure}

\end{document}